\documentclass[a4paper,12pt,fleqn]{article}
\newcommand{\be}{\begin{equation}}
\newcommand{\ee}{\end{equation}}
\newcommand{\bea}{\begin{eqnarray}}
\newcommand{\eea}{\end{eqnarray}}
\begin{document}
%\begin{title}
\centerline {\Large \bf A Class of N-body Problems With Nearest And}
\centerline {\Large \bf Next-to-Nearest Neighbour Interactions}
\vskip 0.3 in
\centerline {Guy Auberson$^{* \dagger}$}
\centerline {Physique Math\'ematique et Th\'eorique, UMR 5825-CNRS,}
\centerline {Universit\'e de Montpellier II, Montpellier, France}

\vskip .3 in

\centerline {Sudhir R. Jain$^{**}$}
\centerline {Theoretical Physics Division, Bhabha Atomic Research Centre,}
\centerline {Trombay, Mumbai 400 085, India}

\vskip .3 in

\centerline {Avinash Khare$^{+ \dagger}$}
\centerline {Institute of Physics, Sachivalaya Marg,}
\centerline {Bhubaneswar 751 005, Orissa, India}
%\maketitle

\vskip .3 in

\begin{abstract}
We obtain the exact ground state and a part of the excitation
spectrum in one dimension
on a line and the exact ground state on a
circle in the case where $N$
particles are interacting via nearest and next-to-nearest
neighbour
interactions.
Further, using the exact ground state,
we establish a mapping between these $N$-body
problems and the
short-range Dyson models introduced recently to model
intermediate
spectral statistics. Using this mapping we compute
the one- and two-point
functions of a related many-body theory
and show the absence of long-range order
in the thermodynamic limit.
However, quite remarkably, we prove the existence of an
off-diagonal long-range order in the symmetrized version of the related
many-body theory.
Generalization of the models
to other root systems is
also considered. Besides, we also generalize the model on the full line to
higher dimensions. Finally, we consider a model in two dimensions in which all
the states exhibit novel correlations.
\vfill


{\bf Classification codes :} 05.30.-d, 05.40.-a, 03.75.Fi

* auberson@lpm.univ-montp2.fr

** srjain@apsara.barc.ernet.in (Corresponding author)

+ khare@iopb.res.in

$\dagger $ {\it Work supported in part by the Indo-French centre for the
promotion of advanced research (CEFIPRA project 1501-1502)}.
\end{abstract}
\newpage
\tableofcontents
\newpage
\section{Introduction}

In recent years, the Calogero-Sutherland  type $N$-body problems
\cite{ca,su} in one dimension have attracted considerable attention
not only because they are exactly solvable \cite{op} but also due to their
relationship with (1+1)-dimensional conformal field theory, random
matrix theory \cite{sla} etc. In particular, the
connections between exactly solvable models \cite{mattis}
and random matrix theory \cite{mehta}
have been very fruitful.
For example,
by mapping these models to random matrices from an orthogonal, unitary or
symplectic Gaussian ensemble,
Sutherland
\cite{su} was able to obtain all static correlation
functions of the
corresponding many body theory. The key point of this model is the
pairwise long-range interaction among the $N$ particles.
One may add here that the family consisting of exactly
solvable models, related to fully integrable systems,
is quite small \cite{op} and their importance lies in the fact that their small
perturbations describe wide range of physically interesting situations.
Further, recent developments \cite{sre}
relating equilibrium
statistical mechanics to random matrix theory owing to non-integrability
of dynamical systems has made the pursuit of unifying seemingly
disparate ideas a very important theme. The results presented in this
paper belong to the emerging intersection of several frontiers like
quantum chaos, random matrix theory, many-body theory and equilibrium
statistical mechanics \cite{jp}.

The universality in level correlations in linear (Gaussian) random matrix
ensembles
agrees very well with those in chaotic quantum systems \cite{bohigas}
as also in many-body systems like nuclei \cite{mehta}. On the other hand, random
matrix theory was connected to the world of exactly solvable models when the
Brownian motion model was presented by Dyson \cite{dyson},
and later on, by the works on level dynamics \cite{haake}. However, there
are dynamical systems which are neither chaotic nor integrable - the so-called
pseudo-integrable systems \cite{jain-lawande}.
It is known that the spectral statistics of such
systems are ``non-universal with a universal trend" \cite{parab-jain}.
In particular, for Aharonov-Bohm billiards, the level spacing distribution
is linear for small spacing and it falls off exponentially for large spacing
\cite{murthy}. Similar features are numerically observed for the Anderson model
in three dimensions at the metal-insulator transition point \cite{guhr}.
To understand these statistical features, and in the context of random banded
matrices, a new random matrix model (which has been called as the short-range
Dyson model in \cite{gremaud}) was introduced \cite{pandey,bogomolny} wherein
the energy levels are treated as in the Coulomb gas model with the difference
that only nearest neighbours interact. This new model explains features
of intermediate statistics \cite{gremaud} in some polygonal billiards.

In view of all this it is worth enquiring if one can construct an $N$-body
problem which is exactly solvable and which is connected to the short-range
Dyson model (SRDM)? If possible, then using this correspondence one can
hope to calculate the correlation functions of the corresponding many-body
theory and see if the system exhibits long-range order and/or off-diagonal
long-range order.

The purpose of
this paper is to present two such models in one dimension, one on a line
and the other on a circle.
We obtain the exact ground state and a part of the excitation spectrum on
a line and the exact ground state on a circle in case the $N$ particles
are interacting via nearest and next-to-nearest neighbour interactions
\cite{jk}.
Further, in both the cases we show how the norm of the ground state
wave function is
related to the joint probability density function of the eigenvalues of
short-range Dyson models. Using this mapping, we obtain one- and
two-point functions
of a related many-body theory in the thermodynamic
limit and prove the absence of long-range order in the system.
However, quite remarkably, we prove the existence of
an off-diagonal long-range
order in the symmetrized version of the corresponding many-body theory
\cite{ajk}.

We also extend this work in several different directions. For example,
we consider an $N$-body problem with nearest and next-to-nearest neighbour
interaction in an arbitrary number of
dimensions $D$ and show that the ground
state and a part of the excitation spectrum can
still be obtained analytically.
We also obtain a part of the bound state spectrum in one dimension (both on
a full line and on a circle)
by replacing the root system $A_{N-1}$ by $BC_N, D_N$ etc. Besides, we also
consider a model in two dimensions for which novel
correlations are present in the ground as well as the excited states.

The plan of the paper is the following. In Sec.II we
consider an $N$-body problem on a line characterized by the Hamiltonian
(throughout this paper we shall use $\hbar =m =1$)
\bea\label{1}
H = - {1\over 2} \sum^N_{i=1} {\partial^2\over \partial x^2_i} +
g \sum^{N-1}_{i=1}
{1\over (x_i-x_{i+1})^2} &-& G \sum^{N-1}_{i=2}
{1\over (x_{i-1}-x_i)(x_i-x_{i+1})}   \nonumber \\
&+& V\left(\sum^N_{i=1} x^2_i\right)
\eea
with $G \ge 0$ while $g > -1/4$ to prevent the collapse that a more attractive
inversely quadratic potential would cause.
We show that the ground state and at least a part of the excitation
spectrum can be obtained if
\be\label{2}
g = \beta (\beta-1), \ G = \beta^2,
V = {\omega^2\over 2} \sum^N_{i=1} x_i^2 \, .
\ee
Note that with the above restriction on $G$ and $g$, $\beta \ge 1/2$.
Further we also point out the connection
between the norm of the ground state wave function and
the joint probability distribution function for eigenvalues in
SRDM.
In Sec. III we consider another
$N$-body problem, but this time on a circle characterized
by the Hamiltonian
\bea\label{3}
&& H =  - {1\over 2} \sum^N_{i=1}{\partial^2\over\partial x^2_i}
+ g{\pi^2\over L^2} \sum^{N}_{i=1}
{1\over \sin^2 [{\pi\over L}(x_i-x_{i+1})]} \nonumber\\
&& - G {\pi^2\over L^2} \sum^{N}_{i=1} \cot \left[ (x_{i-1}
-x_i) {\pi\over L}\right] \cot \left[(x_i-x_{i+1}){\pi\over L} \right]
\, , \ (x_{N+1} = x_1) \, ,
\eea
(where again $G \ge 0$ while $g > -1/4$) and obtain the exact ground state
in case $g$ and $G$ are again as related
by eq. (\ref{2}). Further, we also point out the connection between the
norm of the ground state wave function and the joint probability
distribution function for eigenvalues of short-range circular
Dyson model (SRCDM).
Using this connection, in Secs. IV and V we  obtain
several exact results for the corresponding
many-body theory in the thermodynamic limit. In
particular, in Sec. IV
we calculate the two-particle correlation functions of a related
many-body theory in the thermodynamic limit and prove
the absence of long-range order in the system. In Sec. V we
consider the symmetrized version of the model considered in Sec. III
and show
the existence of an off-diagonal long-range order in the bosonic system in the
thermodynamic limit.
In Sec. VI we consider the $BC_N$ generalization
of the model (1) and obtain the exact ground state of the system.
In Sec. VII we consider the $BC_N$ generalization
of the model (3) and
obtain the exact ground state of the system.
In Sec. VIII we consider a generalization of the model (1) to higher dimensions
and obtain the ground state and a part of the excitation spectrum.
In Sec. IX we consider a
variant of the model (\ref{1}) in two dimensions and obtain the
ground state as well a class of excited states all of which have a novel
correlation built into them.
Finally, in Sec. X
we summarize the results obtained and point out several open questions.

\section{N-body problem in one dimension on a line}

Let us start from the Hamiltonian (\ref{1}) and restrict our attention to the
sector of configuration space corresponding to a definite ordering of the
particles, say
\be\label{5}
x_i \ge x_{i+1} \, , \ \ i = 1,2,...,N-1 \, .
\ee
On using the ansatz
\be\label{2.1}
\psi = \phi \prod^{N-1}_{i=1} (x_i-x_{i+1})^{\beta} \, ,
\ee
in the corresponding Schr\"odinger equation $H\psi = E\psi$, it is easily shown
that, provided $g$ and $G$ are related to $\beta$ by eq. (\ref{2}),
$\phi$ satisfies the equation
\be\label{2.2}
- {1\over 2} \sum^N_{i=1} {\partial^2\phi\over \partial x^2_i} - \beta
\sum^{N-1}_{i=1} {1\over (x_i-x_{i+1})} \left({\partial \phi\over\partial x_i}
-{\partial
\phi\over \partial x_{i+1}}\right) + (V-E)\phi = 0 \, .
\ee
Following Calogero we start from
$\phi$ as given by eq. (\ref{2.2}) and assume that
\be\label{2.3a}
\phi = P_k (x) \Phi (r) \, .
\ee
where $r^2 = \sum^{N}_{i=1} x_i^2$.
The function, $\Phi$ satisfies the
equation
\be\label{2.3b}
\Phi'' (r) + [N+2k-1+2(N-1)\beta]{1\over r}\Phi'(r) +2[E-V(r)]\Phi (r) = 0 \, ,
\ee
provided $P_k (x)$ is
a homogeneous polynomial of degree
$k$ ($k = 0,1,2,...$) in the particle-coordinates and satisfies generalized
Laplace equation
\be\label{2.4a}
 \bigg [ \sum^N_{i=1} {\partial^2 \over \partial x^2_i} + 2\beta
 \sum^{N-1}_{i=1} {1\over (x_i-x_{i+1})} \left({\partial \over\partial x_i}
-{\partial
\over \partial x_{i+1}}\right) \bigg ] P_k(x) = 0 \, .
\ee
We shall discuss few solutions of the Laplace equation (\ref{2.4a}) below.

Let us now specialize to the case of the oscillator potential i.e. $V(r)
= {\omega^2\over 2} r^2$. In this case,  eq. (\ref{2.3b}) is the well known
radial equation for the oscillator problem in
more than one dimension and its solution
is
\be\label{2.5}
\Phi (r) = \exp ({-\omega r^2/2}) L^a_n (\omega r^2), \ n = 0,1,2,....
\ee
where $L^{a}_n(x)$ is the associated Laguerre polynomial while the energy
eigenvalues are given by
\be\label{2.6}
E_{n} = \left[2n+k+{N\over 2} + (N-1)\beta \right] \omega
= E_0 +(2n+k)\omega \, ,
\ee
with $a = {E\over \omega}-2n-1$. Few comments are in order at this stage.
\begin{enumerate}
\item For large $N$, the energy $E$ is proportional to $N$ so that
\be\label{2.7}
\lim_{N\rightarrow\infty} {E\over N}
= \left(\beta +{1\over 2}\right)\omega \, ,
\ee
i.e., the system has a good thermodynamic limit. In contrast, notice that the
long-ranged Calogero model does not have a good thermodynamic limit
since in that
case for large $N$, $E/N$ goes like $N$.
\item  The
spectrum can be interpreted as due to noninteracting bosons
(or fermions) plus $(n,k)$-
independent (but $N$-dependent) shift.
\end{enumerate}

The ground state eigenvalue and eigenfunction of the model is thus given by
($n = k = 0$)
\be\label{2.8}
E_0 = \left[ (N-1)\beta+{N\over 2}\right] \omega \, ,
\ee
\be\label{2.9}
\psi_0 = \exp \left({-{\omega\over 2}\sum^N_{i=1}x^2_i}\right) \prod^{N-1}_{i=1}
(x_i-x_{i+1})^{\beta} \, .
\ee

A neat way of proving that we have indeed obtained the ground state can be
given using the method of supersymmetric quantum mechanics \cite{cks}.
To this end, we define the operators
\bea\label{2.10a}
Q_i & = & {d\over dx_i} +\omega x_i+\beta \bigg [{1\over (x_{i-1}-x_i)}
-{1\over (x_i -x_{i+1})} \bigg ] \, , \ (i = 2,3,...,N-1) \, , \nonumber\\
Q_1 & = & {d\over dx_1} +\omega x_1-\beta {1\over {x_1-x_2}} \, , \nonumber\\
Q_N & = & {d\over dx_N} +\omega x_N+\beta {1\over {x_{N-1}-x_N}} \, ,
\eea
and their Hermitian conjugates $Q_i^{+}$. It is easy to see that the $Q's$
annihilate the ground state as given by eq. (\ref{2.9}). Further, the
Hamiltonian (\ref{1}) can be written in terms of these operators as
\be\label{2.8a}
H - E_0 = {1\over 2}\sum_{i=1}^{N} Q_i^{+} Q_i \, ,
\ee
where $E_0$ is as given by eq. (\ref{2.8}). Now since the operator on the
right hand side is nonnegative and annihilates the ground state wavefunction
as given by eq. (\ref{2.9}), hence $E_0$ as given by eq. (\ref{2.8}) must be
the ground state energy of the system.

On rewriting $\psi_0$ in terms of a new variable
\be\label{2.10}
y_i \equiv \sqrt{{\omega\over\beta}} x_i \, ,
\ee
one finds that the probability distribution for $N$ particles is given by
\be\label{2.11}
\psi^{2}_0 =
C \exp \left(-\beta \sum^N_{i=1} y^2_i\right) \prod^{N-1}_{i=1} (y_i-y_{i+1})^{2\beta}
\ee
where $C$ is
the normalization constant. We now observe that for $\beta = 1,2,4,$
this $\psi^2$ can be identified with the joint probability density
function for the eigenvalues of SRDM with
Gaussian orthogonal, unitary or symplectic ensembles respectively. We
can therefore borrow the well-known results for these ensembles
\cite{bogomolny,pandey}
and obtain exact results about a many-body theory defined in
the limit, $N\rightarrow\infty, \ \omega \rightarrow 0,
\ N \omega =$ finite which
defines the density of the system. For example,
as $N \to \infty $, the one-point function
tends to a Gaussian for any $\beta$
\cite{pandey} and is given by
\be
R_1 (x) = {N \over \sqrt{2\pi \sigma^2}}
\exp{(-{x^2\over 2\sigma^2})} \, ,
\ee
where $\sigma^2 = {(\beta+1) \over \omega}$. Other results about
the many-body theory will be discussed in Secs. IV and V.

Finally, let us discuss the polynomial
solutions to the Laplace equation (\ref{2.4a}).
So far, we have been able to obtain solutions in the following cases:

(i) $k=2, N \ge 2 ~(ii) k=3, N \ge 3 ~(iii) k=4, N \ge 4 ~(iv) k=5, N \ge 5
~(v) k =6, N \ge 6$.

Besides we have also obtained solutions for $k = 4,5,6$ in case $N =3$,
and for $k=5,6$ in case $N=4$. We find that
for $k \ge 3$, the demand that there be no pole in $P_k (x)$
alone does not require $P_k (x)$ to be completely symmetrical polynomial.
However, for $k =3,4$ and $N=3,4$ it turns out that solution to Laplace
eq. (\ref{2.4a}) exists only if $P_k (x)$ is a completely symmetric polynomial.
We suspect that this may be true in general. On assuming completely
symmetric $P_k (x)$ we find that in all the above cases we have a
one-parameter family of solutions. In particular the various solutions are
as follows (it is understood that the particle indices $i,j,k,...$ are
always unequal
unless mentioned otherwise).

(i) $k =2, N \ge 2$

\be
P_k (x) = a \sum_{i=1}^{N} x_i^2 + b \sum_{i<j}^{N} x_i x_j \, ,
\ee
with $\beta$ given by
\be
\beta = \frac{aN}{(N-1)(b-2a)} \, .
\ee

(ii) $k =3, N \ge 3$

\be
P_k (x) = a \sum_{i=1}^{N} x_i^3 + b \sum_{i,j=1}^{N} x^2_i x_j
+c \sum_{i <j <k}^{N} x_i x_j x_k \, ,
\ee
where $c = 3(b-a)$ and
$\beta$ is given by
\be
\beta = \frac{3a+(N-1)b}{(N-1)(b-3a)} \, .
\ee

(iii) $k = 4, N \ge 4$

\bea
&& P_k (x) = a \sum_{i=1}^{N} x_i^4 + b \sum_{i,j=1}^{N} x^3_i x_j
+ c\sum_{i <j}^{N} x_i^2 x_j^2 \nonumber\\
&& +d \sum_{i,j < k}^{N} x_i^2 x_j x_k
+e \sum_{i <j <k <l}^{N} x_i x_j x_k x_l \, ,
\eea
where
\bea
&& e = 6(c-2a) \, , \ d = b+2c-4a \, , \nonumber\\
&& (N+4)b +2(N-2)c -4(N-2)a +2(N-1)(2a+b-c)\beta = 0 \, ,
\eea
and
$\beta$ is given by
\be
\beta = \frac{6a+(N-1)c}{(N-1)(b-4a)} \, .
\ee

(iv) $k = 5, N \ge 5$

\bea
&& P_k (x) = a \sum_{i=1}^{N} x_i^5 + b \sum_{i,j=1}^{N} x^4_i x_j
+ c\sum_{i,j=1}^{N} x_i^3 x_j^2 +d \sum_{i,j < k}^{N} x_i^3 x_j x_k \nonumber\\
&& +e\sum_{k,i<j}^{N} x_i^2 x_j^2 x_k
+f \sum_{i,j<k<l}^{N} x_i^2 x_j x_k x_l
+g \sum_{i <j <k <l < m}^{N} x_i x_j x_k x_l x_m \, ,
\eea
where
\bea
&& e = 5c-5a-3b \, , \ d = b+2c-5a \, , \nonumber\\
&& f = 12c -15a -9b \, , \ g = 30(c-a-b) \, , \nonumber\\
&& (5N-7)c -3(N-4)b -5(N-2)a  \nonumber\\
&& +(N-1)(5a+3b-2c)\beta = 0 \, ,
\eea
and
$\beta$ is given by
\be
\beta = \frac{10a+(N-1)c}{(N-1)(b-5a)} \, .
\ee

(iv) $k = 6, N \ge 6$

\bea
&& P_k (x) = a \sum_{i=1}^{N} x_i^6 + b \sum_{i,j=1}^{N} x^5_i x_j
+ c\sum_{i,j=1}^{N} x_i^4 x_j^2 +d \sum_{i,j < k}^{N} x_i^4 x_j x_k
+e\sum_{i<j}^{N} x_i^3 x_j^3 \nonumber\\
&& +f\sum_{ij,k=1}^{N} x_i^3 x_j^2 x_k
+g \sum_{i,j<k<l}^{N} x_i^3 x_j x_k x_l
+h\sum_{i<j<k}^{N} x_i^2 x_j^2 x_k^2
+p \sum_{i <j, <k <l}^{N} x_i^2 x_j^2 x_k x_l \nonumber\\
&& +q \sum_{i,j <k <l < m}^{N} x_i^2 x_j x_k x_l x_m
+r \sum_{i <j <k <l < m <n }^{N} x_i x_j x_k x_l x_m x_n \, ,
\eea
where
\bea
&& 3e = 4b-2c+6a+f \, , \ d = b+2c-6a \, , g=2f+2c-4b-6a \, , \nonumber\\
&& h = 2f+9a-4b-c \, , p=5f+18a-8b-6c \, , \nonumber\\
&& q=6(2f+9a-4b-3c) \, , \ r= 30(f+6a-2b-2c) \, , \nonumber\\
&& (5N-9)f -2(4N-15)b +18(N-5)a -6(N-5)c \nonumber\\
&& +(N-1)(8b+6c-2f-18a)\beta
= 0 \, , \nonumber\\
&& 14b-2c+6a+(N-1)f+2(N-1)(3a+2b-c)\beta = 0 \, ,
\eea
and
$\beta$ is given by
\be
\beta = \frac{15a+(N-1)c}{(N-1)(b-6a)} \, .
\ee

It would be nice if one can find solutions for higher values of $k$ and
further check if solutions exist (if at all) only if $P_k (x)$
is a completely symmetric polynomial. While we are unable to prove it, we
suspect that, subject to the solutions of the Laplace equation for higher k,
we have obtained the complete spectrum for this problem.

Finally it is worth enquiring if the bound state spectrum of the
Hamiltonian (\ref{1}) can also be obtained in case the oscillator
potential is replaced by any other potential. It turns out that as in the
Calogero case \cite{kh}, in this case also the answer to the question is
in affirmative. In particular, if instead the $N$ particles are interacting
via the $N$-body potential as given by
\be\label{4}
V(x_1,x_2,...,x_N)
= -\alpha \sum^{N}_{i=1} {1 \over \sqrt{\sum_i x_i^2}} ,
\ee
then also (most likely the entire)
discrete spectrum can be obtained. This is because, after using the ansatz
(\ref{2.3a}),
eq. (\ref{2.3b})
is essentially the radial Schr\"odinger equation for an
attractive Coulomb potential and it is well known that
the only two problems which
are analytically solvable for all partial waves are the Coulomb and the
oscillator potentials. In particular the solution of  (\ref{2.3b})
is then given
by (note $r^2 = \sum^N_{i=1} x^2_i$)
\be\label{2.14}
\Phi(r) = \exp (-\sqrt{2\mid E\mid} r) L^b_n (2\sqrt{2\mid E\mid} r) \, ,
\ee
and the corresponding energy eigenvalues are
\be\label{2.15}
E_{n,k} = - {\alpha^2\over 2 \bigg [n+k+{N\over 2}-1+(N-1)\beta \bigg ]^2} \, ,
\ee
when $ b = N+2k-3+2(N-1)\beta$. It may be noted that whereas in the
oscillator case the
spectrum is linear in $\beta$, it is $(-E)^{-1/2}$ which is
linear in $\beta$ in the case of the Coulomb-like potential. Secondly, as
in any oscillator (Coulomb) problem, the energy depends on $n$ and $k$
only through
the combination $2n+k$ ($n+k$).

Is there
any underlying reason why one is able to obtain the discrete spectrum
for the $N$-body problem
with either the oscillator or the Coulomb-like potential
(\ref{4}) ? Following \cite{gk}
it is easily shown that in both the cases one can
write down an underlying $SU(1,1)$ algebra. Further, since the many-body
potential $W$ in  (\ref{1}) is a
homogeneous function of the coordinates of degree
-2, i.e. it satisfies
\be\label{2.20}
\sum^N_{l=1} x_l {\partial W\over\partial x_l} = - 2W \, ,
\ee
hence, following the arguments of \cite{gk}, one can also establish a simple
algebraic relationship between the energy eigenstates of the $N$-body
problem (\ref{1}) with the Coulomb-like potential (\ref{4}) and
the harmonic oscillator potential.

It may be noted that the Hamiltonian (\ref{1}) is not completely symmetric
in the sense that whereas all other particles have two neighbours, particle
1 and $N$ have only one neighbour. Can one make it symmetric so that all
particles will be treated on the same footing? One possible way is to add
some extra terms in $H$. For example, consider
\be\label{21}
H_1 = H + H' \, ,
\ee
where $H$ is as given by eq. (\ref{1}) while $H'$ has the form
\be\label{22}
H' = \frac{g}{(x_N - x_1)^2} -G \bigg [\frac{1}{(x_N -x_1)(x_1-x_2)}
+\frac{1}{(x_{N-1} -x_N)(x_N -x_1)} \bigg ] \, .
\ee
Clearly, by adding these extra terms, the problem has become cyclic invariant
for any $N$ while for $N=3$ it is identical to the Calogero problem and hence
is in  fact completely symmetric under the interchange of any two of the three
particle coordinates.
It may be noted that in the thermodynamic limit, these extra terms are
irrelevant.

We can again start from the ansatz (\ref{2.1}) (but with $N-1$ replaced by $N$)
in the Schr\"odinger equation $H_1 \psi = E\psi$ and using eq. (\ref{2}) we
find that $\phi$ again satisfies eq. (\ref{2.2}) but with $N-1$ in the
second term
being replaced by $N$. On further using the substitution as given by
eq. (\ref{2.3a}) one finds that $\Phi$ satisfies eq. (\ref{2.3b}) but with the
coefficient of the $2\beta$ term being $N$ instead of $N-1$ while $P_k (x)$
is again a homogeneous polynomial of degree $k$ (k=0,1,2,...) in the particle
coordinates, which now satisfies instead of eq. (\ref{2.4a})
\be\label{23}
 \bigg [ \sum^N_{i=1} {\partial^2 \over \partial x^2_i} + 2\beta
 \sum^{N}_{i=1} {1\over (x_i-x_{i+1})} \left({\partial \over\partial x_i}
-{\partial
\over \partial x_{i+1}}\right) \bigg ] P_k(x) = 0 \, ,
\ee
with $x_{N+1} = x_1$.

How do the solutions to the Laplace eqs. (\ref{2.4a}) and (\ref{23}) compare?
For $N=3$, eq. (\ref{23}) is identical to that of Calogero and
for this case Calogero has already obtained the solutions for any k. For $N>3$
and for $k \ge 3$, the demand that there be no pole in $P_k (x)$
alone does not require $P_k (x)$ to be completely symmetrical polynomial.
However, for $k =3,4$ and $N=4$ it again turns out that solution to Laplace
eq. (\ref{23}) exists only if $P_k (x)$ is a completely symmetric polynomial.
We suspect that this may be true in general. On assuming completely
symmetric $P_k (x)$ we
have been able to obtain a
two-parameter family of solutions in case
k=3,4,5,6 and $N \ge k$
(note that for eq. (\ref{2.4a}) we have obtained only one-parameter family of
solutions). As an illustration, the solution for $N \ge 4$ and $k=4$ is
given by
(it is understood that the particle indices $i,j,k,...$ are
always unequal)
\bea
&& P_k (x) = a \sum_{i=1}^{N} x_i^4 + b \sum_{i,j=1}^{N} x^3_i x_j
+ c\sum_{i <j}^{N} x_i^2 x_j^2 \nonumber\\
&& +d \sum_{i,j < k}^{N} x_i^2 x_j x_k
+e \sum_{i <j <k <l}^{N} x_i x_j x_k x_l \, ,
\eea
where
\be
 e = 2(2a+2d-2b-c) \, ,
\ee
\be
6a+(N-1)c +\beta[8a-2b+(2c-d)(N-2)] = 0 \, ,
\ee
\be
6b+(N-2)d +2\beta[2(N-1)b-2(N-4)a+(N-4)c-2(N-1)d] = 0 \, .
\ee

Solution to the new
$\Phi$ equation can be easily written down in case
$V(r) = \frac{\omega^2 r^2}{2}$ or if it is given by eq. (\ref{4}).
For example, it is easily checked that in the former case the solution is
again given by eq. (\ref{2.6}) but the energy eigenvalues are now given by
\be\label{24}
E_{n,k} = [2n+k+\frac{N}{2}+N\beta]\omega = E_0 +(2n+k)\omega \, .
\ee
Similarly, in the later case, the solution is as given by eq. (\ref{2.15})
except that
in the term containing $\beta$, $N-1$ must be replaced by $N$.

Apart from these two potentials, where we have obtained
the entire bound state spectrum,
there are several other potentials which are quasi-exactly solvable.
For example, for the potential
\be\label{2.16}
V \left(\sum x^2_i\right) = A \sum^N_{i=1} x^2_i - B \left(\sum^N_{i=1} x^2_i
\right)^2 +
C \left(\sum^N_{i=1} x^2_i\right)^3 \, ,
\ee
it is easily shown that the ground state energy and eigenfunctions are
\be\label{2.17}
E = -{B\over 4\sqrt C} [N+2(N-1)\beta ] \, ,
\ee
\be\label{2.18}
\psi_0 = \exp \bigg [ -{\sqrt C\over 4} (\sum^N_{i=1} x^2_i)^2
+ {B\over 4\sqrt C}
\sum^N_{i=1} x^2_i \bigg ] \prod^{N-1}_{i=1} (x_i-x_{i+1})^{\beta} \, ,
\ee
provided $A,B,C$ are related by
\be\label{2.19}
A = {B^2\over 4C} - [N+2+2(N-1)\beta ] \sqrt C \, .
\ee
It is worth enquiring if the probability distribution
for $N$ particles corresponding
to (\ref{2.18}) can be mapped to some matrix model. In this context
let us point
out that the corresponding (long-ranged) Calogero problem
was in fact mapped to the
matrix model corresponding to branched polymers \cite{jk1}. So far as we are
aware of, answer to this question is not known in our case.

\section{N-body problem in one-dimension with periodic boundary condition}

Soon after the seminal papers of Calogero \cite{ca} and Sutherland \cite{su}
where they considered an $N$-body problem on full line,
Sutherland \cite{sut} also considered
an $N$-body problem with long-ranged interaction and with periodic boundary
condition. He obtained the exact ground state energy and showed that the
corresponding $N$-particle probability density function is related to the
random matrix with circular ensemble \cite{sut}. Using the known results
for the random matrix theory \cite{mehta}, he was able to obtain the
static correlation functions of the corresponding many body theory.
It is then natural to
enquire if one can also obtain the exact ground state
of an $N$-body problem with
nearest and next-to-nearest neighbour
interaction  with periodic boundary condition (PBC). Further, one would like to
enquire if the corresponding $N$-particle probability density can be mapped
to some known matrix model.
The hope is that in this case one may be able to obtain
the correlation functions
of a related many-body theory in the thermodynamic limit.
 We now show that the answer to the question is
in the affirmative.

Let us start from the Hamiltonian (\ref{3}).
We wish to find the ground state of the system subject to the periodic boundary
condition (PBC)
\be\label{6.2}
\psi(x_1,...,x_i+L, ..., x_N) = \psi (x_1,..., x_i,..., x_N).
\ee
For this, we start with a trial wave function of the form
\be\label{6.3}
\Psi_0 = \prod^{N}_{i=1} \sin^{\beta} \left[{\pi\over L} (x_i-x_{i+1})\right]
\, , (x_{N+1} = x_1) \, .
\ee
In this section, we restrict the coordinates $x_i$ to the sector $L \ge x_1
\ge x_2 \ge ... \ge x_N \ge 0$, so that eq. (\ref{6.3}) makes sense even for
noninteger $\beta$. The extension to the full configuration space will be made
in Sec. 5. On
substituting eq. (\ref{6.3}) in
the Schr\"odinger equation for the Hamiltonian (\ref{3}), we find that
it is indeed a solution provided $g$ and $G$ are again related to $\beta$
by eq. (\ref{2}).
The corresponding ground state energy turns out to be
\be\label{6.5}
E_0 = {N \beta^2\pi^2\over L^2} \, .
\ee

The fact that this is indeed the ground state energy can be neatly proved
by using the operators \cite{gks}
\be\label{6.5a}
Q_i = {d\over dx_i} +\beta {\pi\over L} \bigg [\cot (x_{i-1} -x_i) -
\cot (x_i - x_{i+1}) \bigg ] \, ,
\ee
and their Hermitian conjugates $Q_i^{+}$. It is easy to see that the $Q's$
annihilate the ground state as given by eq. (\ref{6.3}). The Hamiltonian
(\ref{3}) can be rewritten in terms of these operators as
\be\label{6.5b}
H - E_0 = {1\over 2} \sum_i Q_i^{+} Q_i \, ,
\ee
where $E_0$ is as given by eq. (\ref{6.5}).
Hence $E_0$ must be the ground state energy of the system.

Thus unlike the Calogero-Sutherland type of models, our models
(both of Sec. II and here) have good thermodynamic limit, i.e., the ground
state energy per particle $(=E_0/N)$ is finite as $N \rightarrow \infty$.

Having obtained the exact ground state, it is natural to enquire if
the corresponding $N$-particle probability density can be mapped to the joint
probability distribution of some SRCDM so
that we can obtain some exact results for the
corresponding many-body theory.
It turns out that indeed the square of the ground-state wave
function is related to the joint probability distribution
function for the SRCDM from where we conclude that the density
is a constant if $0 \leq x \leq {N \over L}$, and zero outside.
Other exact results for the many-body theory will be
discussed in the next two sections.

\section{Some exact results for the many-body problem}

The square of the ground-state wavefunction of the many-body problem
introduced in Sec.II (Sec.III)
can be identified with the joint probability distribution
function of eigenvalues of the SRDM (SRCDM).
Using SRCDM, Pandey \cite{pandey} and Bogomolny et al. \cite{bogomolny}
have  shown that
 for any $\beta $, the two-point correlation function has the form
\be\label{8.1}
R_2^{(\beta )}(s) = \sum_{n=1}^{\infty} P^{(\beta )}(n,s) \, ,
\ee
where $s$ is the separation of two levels (or distance between two
particles in the many-body theory considered here) and
\be\label{8.2}
P^{(\beta )}(n,s) = \frac{(\beta +1)^{n(\beta + 1)}}{\Gamma [n(\beta + 1)]}
s^{n(\beta + 1) - 1}e^{-(\beta + 1)s} \, .
\ee
>From this expression it is not very easy to compute $R_2(s)$ for arbitrary
$\beta$. However, it is easy to obtain the Laplace transform of $R_2(s)$
for any $\beta$. In particular, if
\be\label{8.2a}
g_2(t) = \int_{0}^{\infty} R_2 (s) e^{-ts} ds \, ,
\ee
then
\be\label{8.2b}
g_2(t) = \sum^{\infty}_{n=1} g(n,t) \, ,
\ee
where $g(n,t)$ is the Laplace transform of $P(n,s)$, i.e.,
\be\label{8.2c}
g(n,t) = \int_{0}^{\infty} P(n,s) e^{-ts} ds \, .
\ee
On using $P^{(\beta)} (n,s)$ as given by eq. (\ref{8.2}) in eq. (\ref{8.2c})
it is easily shown that
\be\label{8.2d}
g^{(\beta)} (n,t) = \left(\frac{\beta +1}{t+\beta +1}\right)^{(\beta+1)n} \, .
\ee
Hence
\be\label{8.2e}
g_2^{(\beta)} (t) = \sum^{\infty}_{n=1} g^{(\beta)} (n,t)
= \frac{1}{(\frac{t+\beta+1}{\beta+1})^{\beta+1} -1} \, ,
\ee
from which one has to compute $R_2^{(\beta)} (s)$ by the Laplace inversion.

For integer $\beta $, it is possible to perform the Laplace inversion by
making use of the fact that
\be\label{8.2f}
\frac{1}{x^n -1} = \frac{1}{n} \sum_{k=0}^{n-1}
\frac{e^{2ik\pi /n}}{x-e^{2ik\pi /n}} \, ,
\ee
yielding
\be\label{8.3}
R_2^{(\beta )}(s) = \sum_{k=0}^{\beta} \Omega^k e^{(\beta + 1)s(\Omega^k - 1)}
\ee
where
\be\label{8.7a}
\Omega = e^{2\pi i/(\beta + 1)} \, .
\ee
For $\beta = 1$, which corresponds to
the orthogonal ensemble, the result is already known \cite{pandey,bogomolny} :
$R_2^{(1)}(s) = 1 - e^{-4s}$.

It is interesting to mention that
$R_2^{(1)}(s)$ agrees very well with  some of the pseudo-integrable billiards
(e.g., the $\frac{\pi}{3}$-rhombus billiard). It is important here to note
that for rhombus billiards \cite{gremaud}, the Hamiltonian matrix has elements
which fall in their magnitude away from the principal diagonal. Thus,
beyond a certain bandwidth, the elements are insignificant and the matrix
is effectively banded. Immediately then, the results of banded matrices
become applicable. Although there seems to be good agreement of the results
from this random matrix theory as shown in \cite{gremaud,bogomolny},
in \cite{gremaud} it
is also shown that there are other polygonal billiards for which $R_2^{(1)}(s)$
is not an appropriate correlator. It is possible that for different bandwidths,
and, by an inclusion of interactions beyond nearest neighbours in the
short-range Dyson model, a family of random matrices result. This may,
eventually, explain the entire family of systems exhibiting intermediate
spectral statistics.

Coming back to the two-point correlation function,
depending on if $\beta$ is an odd or an even integer, $R_2(s)$, as given
by eq. (\ref{8.3}), can be written
in a closed form which shows that $R_2 (s)$ is indeed real and further, it
clearly exhibits oscillations for large $s$.
In particular, it is easily shown that
\bea\label{8.4}
R_2 (\beta = 2p+1,s) & = & 1- e^{-2(2p+2)s}
            +  2e^{-(2p+2)s}
\sum_{m=1}^p e^{(2p+2)s \cos ({m\pi\over p+1})} \nonumber\\
                     &   &\cos \bigg [{m\pi\over p+1}
+(2p+2)s \sin ({m\pi\over p+1}) \bigg ] \, ,
\eea
\bea\label{8.5}
R_2 (\beta = 2p,s) & = & 1
+2e^{-(2p+1)s} \sum_{m=1}^p e^{(2p+1)s \cos ({2m\pi\over 2p+1})} \nonumber\\
                   &   & \cos \bigg [{2m\pi\over 2p+1}+(2p+1)s
\sin ({2m\pi\over 2p+1}) \bigg ] \, .
\eea
For illustration, we give below
explicit expressions for $\beta = 2, 3, 4$
\bea
R_2^{(2)}(s) &=& 1 - 2e^{-\frac{9s}{2}}\cos \left( \frac{3\sqrt{3}s}{2}-
\frac{\pi}{3}\right) ;\nonumber \\
R_2^{(3)}(s) &=& 1 - e^{-8s} - 2e^{-4s}\sin (4s) ;\nonumber \\
R_2^{(4)}(s) &=& 1 + 2e^{5s(-1+\cos (2\pi /5))}\cos\left[\frac{2\pi}{5}
+ 5s\sin \left(\frac{2\pi }{5}\right)\right]\nonumber \\
&+& 2e^{5s(-1+\cos (4\pi /5))}\cos\left[\frac{4\pi}{5}
+ 5s\sin \left(\frac{4\pi }{5}\right)\right].
\eea

In Fig. 1, we have plotted $R^{(\beta)}_2 (s)$ as a function of $s$ for
$\beta = 1,2,3,4$.
These results show that, for integer $\beta $, there is no long-range order
in the corresponding many-body theory.

Similarly, if $\beta$ is half-integral, i.e., $\beta = (2n+1)/2$ then it is
easily shown that
\be
R_2^{((2n+1)/2)} (s) = \frac{1}{2} \sum_{k=0}^{2n} \Omega^{2k}
e^{-\frac{2n+1}{2} s(1-\Omega^{2k})}
\bigg [1+{\rm erf} \bigg (\sqrt{\frac{(2n+1)s}{2}}
\Omega^k \bigg ) \bigg ] \, ,
\ee
where $\Omega$ is as given by eq. (\ref{8.7a}).

For arbitrary $\beta $, however, we are unable to perform the Laplace
inversion and hence we do not have a closed expression for $R_2 (s)$.
However, one can numerically calculate it by using eqs. (\ref{8.1}) and
(\ref{8.2}). In Fig. 2, we have plotted $R_2^{(\beta)} (s)$ as a function
of $s$ for $\beta = 1, 4/3, 3/2, 5/3, 2, 7/3, 5/2$. From this figure it is
clear that even for fractional $\beta$, there is no long-range order.

\section {Off-diagonal long-range order}

So far, nothing has been specified regarding the statistical character
of the particles involved in the N-body problem of Sec. III. We now do
that by first symmetrizing the Hamiltonian, that is by rewriting it as
\bea\label{5.1}
H & = & - \frac{1}{2} \sum^N_{i=1}\frac{\partial^2}{\partial
  x_i^2} \nonumber \\
  & + & \sum_{P \varepsilon S_N}\Theta (x_{P(1)}-x_{P(2)})... \Theta
  (x_{P(N-1)}-x_{P(N)}) W (x_{P(1)},...,x_{P(N)}) \, ,
\eea
where $\theta$ is the step function and $W(x_1,...,x_N)$ is the
N-body potential of eq. (3). Next, relying on the solution given in
eq. (\ref{6.3}), we introduce the (not normalized) wave function:
\be\label{5.2}
\psi_N(x_1,...,x_N) = \varepsilon_P \phi_N (x_{P(1)},... , x_{P(N)}) \, ,
\ee
where P is the permutation in $S_N$ such that $1> x_{P(1)} > x_{P(2)} >... >
  x_{P(N)} > 0, \varepsilon_P = 1 (\varepsilon_P = sign (P))$
in the N-boson (N-fermion)
  case and
\be\label{5.3}
\phi_N (x_1,...,x_N) = \prod^N_{n=1} \mid sin \pi (x_n-x_{n+1})\mid^{\beta} \, ;
 \ \ (x_{N+1}=x_1) \, ,
\ee
(we have set the scale factor $L$ equal to 1). Primitively, the
function (\ref{5.2}) is defined on the hypercube $[0,1]^N$. The following
properties of $\psi_N$ are easily verified, provided that $\beta \ge 2$:

\begin{enumerate}

\item In the bosonic case, $\psi_N$ can be continued to a
  multi-periodic function in the whole space ${\cal{R}}^{N}$ (or
  equivalently on the torus $T^N$):
\be\label{5.4}
\psi_N(x_1,..., x_{i}+1,...,x_N) = \psi_N (x_1,...,x_i,...,x_N) \, ; \ \
(i = 1,...,N) \, ,
\ee
which belongs to $C^2$ (i.e. is twice continuously
differentiable). Owing to this property and the results of Sec.3,
$\psi_N$ then obeys the Schr\"odinger equation (with Hamiltonian (3) and
energy as given by eq. (\ref{6.5})) not only in the sector $x_1 > x_2 > ...
> x_N$ but
every where. Thus, $\psi_N$ describes the ground state wave function
of the N-boson system. Moreover, it is translation invariant (on
${\cal{R}}^{N}$):
\be\label{5.5}
\psi_N (x_1+a,x_2+a,..., x_N+a) = \psi_N (x_1,x_2,...,x_N) \, ; \ \
V \ a \ \varepsilon \ \cal{R} \, .
\ee

\item In the fermionic case, the continuation by periodicity is possible
only for {\bf odd} $N$, in which case eq. (\ref{5.4}) still holds with
$\psi_N \ \varepsilon \ C^2$. For even $N$ on the contrary, enforcing the
periodicity (\ref{5.4}) leads to a discontinuous function $\psi_N$, so that
the Schr\"odinger equation is no longer satisfied on the configuration
space $T^N$.

Therefore, in the following we shall implicitly restrict ourselves to
odd values of $N$ when treating fermions. The translation invariance
(\ref{5.5}) then remains valid.

\end{enumerate}

We are interested in the one-particle reduced density matrix, given by
\be\label{5.6}
\rho_{N} (x-x') = \frac{N}{C_N}\int^1_0 dx_1...\int^1_0 dx_{N-1}
\psi_N (x_1,..., x_{N-1}, x) \psi_N (x_1,...,x_{N-1}, x') \, ,
\ee
where $C_N$ stands for the squared norm of the wave function:
\be\label{5.7}
C_N= \int^1_0 dx_1 ...\int^1_0 dx_N \mid \psi_N (x_1,...,x_N)\mid^2 \, .
\ee
That the R.H.S. of eq. (\ref{5.6}) defines a (periodic) function of $(x-x')$
is an easy consequence of eqs. (\ref{5.4}) and (\ref{5.5}). The normalization
of $\rho_N$ is such that $\rho_N (0)=N$, the particle density. Further, the
function $\rho_N (\xi)$ is manifestly of positive type on the $U(1)$ group,
which implies that its Fourier coefficients,
\be\label{5.8}
\rho^{(n)}_N = \int^1_0 d\xi e^{-2i\pi n\xi} \rho_N(\xi) \, ; \ \
(n=0, \pm 1,\pm 2,...) \, ,
\ee
are non-negative (Bochner's theorem). In fact, this directly appears if one
writes their explicit expression
\bea\label{5.9}
\rho^{(n)}_N & = & \frac{N}{C_N}\int^1_0 dx_{1}...\int^1_0 dx_{N-1}
\psi_N (x_1,..., x_{N-1},0) \ X \nonumber \\
             & X & \int^1_0 dx e^{2 i\pi n x}
\psi_N (x_1,...,x_{N-1},x) \, ,
\eea
in the form (obtained by using the periodicity property):
\be\label{5.10}
\rho^{(n)}_N = \frac{N}{C_N}\int^1_0 dx_1 ... \int^1_0 dx_{N-1}
\mid  \int^1_0 dx e^{2 i\pi n x}\psi_N(x_1,...,x_{N-1},x) \mid^2 \, .
\ee
In the {\bf bosonic} case, since the function $\rho_N$ is not only of
positive type but also {\bf positive} (like $\psi_N$), eq. (\ref{5.8})
shows us that
\be\label{5.11}
\rho^{(0)}_N \geq \rho^{(n)}_N \, ; \ \ (n = \pm 1,\pm 2,...) \, .
\ee
In the fermionic case, eq. (\ref{5.11}) is not necessarily true
(because $\psi_N$ changes sign on $T^N$) and it is not an easy matter
to determine the largest Fourier coefficient. Notice that the coefficients
$\rho_N^{(n)}$, which physically represent the expectation values of the
number of particles having momentum $k_n = 2\pi n$ in the ground state, are
nothing but the eigenvalues of the one-particle reduced density matrix
(diagonal in the $k_n$ representation). According to the Onsager-Penrose
criterion \cite{penrose-onsager},
no condensation can occur in the system (at least for Bose
particles) if the largest of these eigenvalues is not an extensive quantity
in the thermodynamic limit, that is, if
\be\label{5.12}
\lim_{N\rightarrow\infty} \frac{\rho^{(0)}_N}{N} = 0 \, .
\ee
For Fermi particles, this criterion is not sufficient, and one has  to look
also at the largest eigenvalue of the two-particle reduced density
matrix \cite{yang}. Since we are presently unable to determine the largest
eigenvalue of $\rho_N$ itself in the fermionic case, we shall not
discuss extensively the latter here. Nevertheless, we shall look for
the large $N$ behaviour of $\rho^{(0)}_N$ for bosons and fermions at a
time, as this does not require much extra work and can give some
indications in the fermionic case too. Let us write:
\be\label{5.13}
\frac{\rho^{(0)}_N}{N} = \frac{A_N}{C_N} \, ,
\ee
where $C_N$ is given by eq. (\ref{5.7}) and
\be\label{5.14}
A_N = \int^1_0 dx_1...\int^1_0 dx_{N-1} \psi_N (x_1,...,x_{N-1},
0) \int^1_0 dx \psi_N (x_1,...,x_{N-1}, x) \, ,
\ee
(the expression (\ref{5.9}) of $\rho^{(0)}_N$ is more convenient than
(\ref{5.10}) for
our purpose). Because of the special form (\ref{5.2})-(\ref{5.3}) of the wave
function, the computation of the squared norm $C_N$ is already not a
trivial task, in sharp contrast to the case of N free, impenetrable
particles. As a consequence, the (mainly algebraic) method introduced
long ago by Lenard \cite{lenard} to deal with the latter case does not apply
here, and we have to resort to another device. For conciseness, we
introduce the notation:
\be\label{5.15}
S (x_n - x_{n-1}) _ = \mid \sin \pi (x_n- x_{n+1})\mid^{\beta} \, ,
\ee
and define:
\be\label{5.16}
 S_2 (\triangle) = \int^{\triangle}_0 dx S(x) S(\triangle-x) \, ; \ \ (0\leq
\triangle\leq 1) \, .
\ee
Our starting point will be the following representations of $C_N$ and
$A_N$:
\be\label{5.17}
C_N = (N-1)! \frac{1}{2\pi} \int^{\infty}_{-\infty} dx e^{-ix}\tilde
{F} (x)^N \, ,
\ee
\be\label{5.18}
 A_N = (N-1)! \frac{1}{2\pi} \int^{\infty}_{-\infty} dx e^{-ix}\tilde
{F}(x)^{N-3} \bigg [\tilde {F}(x)\tilde {G}(x)
+\eta_N\tilde {H}(x)^2 \bigg ] \, ,
\ee
where
\bea\label{5.19}
\tilde{F} (x) & = & \int^1_0 d\triangle e^{i\triangle x} S(\triangle)^2 \, ,
\nonumber \\
\tilde{G} (x) & = & \int^1_0 d\triangle e^{i\triangle x} S_2(\triangle)^2 \, ,
\nonumber \\
\tilde{H} (x) & = & \int^1_0 d\triangle e^{i\triangle x} S(\triangle)
S_2 (\triangle)^2 \, ,
\eea
and
\be\label{5.20}
\eta_{N} =  \begin{array}{ll}
(N-2) &\mbox{for bosons} \\
-1 &\mbox {for fermions} \, .
\end{array}
\ee
The representations (\ref{5.17})-(\ref{5.20}) follow from the
convolution structure of
the expressions (\ref{5.7}) and (\ref{5.14}) of $C_N$ and $A_N$, when
written in terms
of appropriate variables. Their proof is given in the Appendix. Our
aim is to extract from them the large $N$ behaviour of $C_N$ and
$A_N$. Their form is especially suited for that purpose, because  the
integrands in eqs. (\ref{5.17}) and (\ref{5.18}) are entire functions,
as polynomial
combinations of Fourier transforms of functions with compact support
(eq. (\ref{5.19})). Indeed, we are then allowed to, first, shift the
integration path
and then apply the residue theorem to meromorphic pieces of the
integrands. However, it turns out that the calculations needed for
arbitrary (integer) values of $\beta$ are quite cumbersome. So, in
order to keep the argument clear enough, we shall content ourselves to
present below these calculations in the simplest case, namely $\beta$
= 1 (recall that, strictly speaking, this value is not allowed), being
understood that similar results are obtained for all integers $\beta
\ge 2$.

For $\beta = 1, S(\triangle) = \sin \pi\triangle$, and eq. (\ref{5.19})
gives, after reductions:
\bea\label{5.21}
\tilde {F} (x) & = & \frac{2\pi^2}{i}\frac{1-e^{ix}}{x(x^2-4\pi^2)} \, ,
\nonumber \\
\tilde {G} (x) & = & \frac {4\pi^4}{i} \frac{5x^2-4\pi^2}{x^3(x^2-4\pi^2)^3}
+ e^{ix} R^{(-1)}(x) \, , \nonumber \\
\tilde {H} (x) & = & -\frac {4\pi^3}{i}\frac {1}{x(x^2-4\pi^2)^2}
+ e^{ix} R^{(-2)}(x) \, ,
\eea
where $R^{(n)}(x)$ is a generic notation for rational functions
behaving like $x^n$ when $x\rightarrow \infty$, and the precise form
of which will be eventually of no importance. This produces, for the
functions to be integrated in eqs. (\ref{5.17}) and (\ref{5.18}):
\be\label{5.22}
\tilde {F} (x)^N = (\frac{2\pi^2}{i})^{N} \bigg [\frac {1}{[x(x^2-4\pi^2)]^N} +
\sum^N_{n=1} e^{inx} R_n^{(-3N)} (x) \bigg ] \, ,
\ee
\bea\label{5.23}
&& \tilde {F} (x)^{N-3} [ \tilde {F} (x)\tilde {G} (x)
+\eta_N \tilde {H} (x)^2 ] \nonumber \\
&& = i
\left(\frac {2\pi^2}{i}\right)^N \bigg \{\frac {(5+2\eta_N)x^2-4\pi^2}
{[x(x^2-4\pi^2)]^{N+1}}
+ \sum^{N+1}_{n=1} e^{inx} R_n^{(-3N-1)} (x) \bigg \}
\, .
\eea
Let us stress again that these functions, when analytically continued,
are holomorphic in the whole complex plane (the poles appearing in the
first term are exactly canceled by the remaining ones).

We consider first $C_N$, now given by
\be\label{5.24}
C_N = (N-1)! \left(\frac{2\pi^2}{i}\right)^N \frac {1}{2\pi}
\int^{\infty}_{-\infty} dx e^{-ix} \bigg \{ \frac{ 1}{[x(x^2-4\pi^2)]^N} +
\sum^N_{n=1} e^{inx} R_n^{(-3N)} (x) \bigg \} \, .
\ee
Since the function within the curly bracket is an entire one, we
can shift the integration path to $I \equiv \{ z = x+ia \ \
\mid x \ \varepsilon \ \cal{R} \}$.
Let us choose $a > 0$. Then, by Cauchy theorem
\be
\int_I dz e^{-iz}\sum^N_{n=1} e^{inz} R_n^{(-3N)} (z) = 0 \, .
\ee
Indeed, the integrand is holomorphic above $I$ and is bounded there by
const. $\mid z \mid^{-3N}$, which allows us to close the integration
path at infinity in the {\bf upper} complex plane. We end up with
\be\label{5.25}
C_N= (N-1)! \left(\frac{2\pi^2}{i}\right)^N \frac {1}{2\pi}\int_I dz \frac
{e^{-iz}} {z^N(z^2-4\pi^2)^N} \, .
\ee
Similarly, we are allowed to close the integration path at infinity in
eq. (\ref{5.25}), but this time in the {\bf lower} complex plane.
The integrand has
now poles at $z = 0, \pm 2\pi$, and applying the residue theorem
leads to explicit expressions for $C_N$. Unfortunately, these
expressions turn out to appear as (finite) sums with alternating
signs, the terms of which become very close to each other for large
$N$. They are therefore useless for determining the asymptotic behaviour
of $C_N$, and we have to proceed differently. Let us write
\bea\label{5.26}
& \int_I dz \frac{e^{-iz}}{z^N(z^2-4\pi^2)^N} = \frac {1} {(N-1)!}
\frac {d^{N-1}}{d\alpha^{N-1}}\mid_{\alpha=4\pi^2} \int_I dz
\frac {e^{-iz}}{z^N(z^2-\alpha)} \nonumber \\
& = \frac {-2i\pi} {(N-1)!}
\frac {d^{N-1}}{d\alpha^{N-1}}\mid_{\alpha=4\pi^2} [
R_{+} (\alpha) +R_{-} (\alpha) +R_{0} (\alpha)] \, ,
\eea
where $R_{\pm} (\alpha)$  and $R_0 (\alpha)$ are the residues of the last
integrand at $z = \pm \sqrt{\alpha} $ and $ z = 0$ respectively. They
are readily computed, assuming first that $N=2M+1$ is odd:
\bea\label{5.27}
R_{+} (\alpha) +R_{-} (\alpha) & = & \frac {\cos \sqrt{\alpha}}
{\alpha^{M+1}} = \sum^{\infty}_{r=0}
\frac{(-1)^r}{(2r)!}\alpha^{r-M-1} \, , \nonumber \\
               R_{0} (\alpha)  & = & -  \sum^{M}_{r=0}
\frac{(-1)^r}{(2r)!}\alpha^{r-M-1} \, .
\eea
Hence
\be\label{5.28}
R_{+}(\alpha)+R_{-}(\alpha)+R_{0}(\alpha) = (-1)^{M+1}\sum^{\infty}_{s=0}
\frac {(-1)^s} {(2M+2s+2)!} \alpha^s
\ee
Using  eqs. (\ref{5.25}), (\ref{5.26}) and (\ref{5.28}) we then obtain
\bea\label{5.29}
C_N & = & \left(\frac {2\pi^2}{i}\right)^N (-1)^{M+1}(-i)
\frac{d^{N-1}}{d\alpha^{N-1}}\mid_{\alpha=4\pi^2} \sum^{\infty}_{s=0}\frac
{(-1)^s}{(2M+2s+2)!}\alpha^s \nonumber \\
    & = & (2\pi^2)^N\sum^{\infty}_{n=0} \frac{(N+n-1)!}{n!(3N+2n-1)!}
(-4\pi^2)^n \, .
\eea
The result is exactly the same for even $N$. It suffices now to observe
that the last series alternates in sign and is decreasing to
deduce
\be\label{5.30}
C_N = (2\pi^2)^N \frac{(N-1)!}{(3N-1)!} [ 1+O(\frac{1}{N})] \, .
\ee

Our procedure for evaluating $A_N$ is quite similar, and we give below
only the main steps. From eqs. (\ref{5.18}) and (\ref{5.23}) we get
\bea\label{5.31}
& A_N = (N-1)! (\frac {2\pi^2}{i})^N \frac {i}{2\pi}\int_I dz e^{-iz}
\frac {(5+2\eta_N)z^2 - 4\pi^2} {Z^{N+1}
  (z^2-4\pi^2)^{N+1}} \nonumber \\
& =  \frac {1}{N} (\frac{2\pi^2}{i})^N \frac{i}{2\pi}
\frac {d^N}{d\alpha^N}\mid_{\alpha=4\pi^2} \int_I dz e^{-iz}
\bigg [\frac{5+2\eta_N} {z^{N-1}(z^2-\alpha)}
- \frac {4\pi^2}{z^{N+1}(z^2-\alpha)} \bigg ] \, ,
\eea
and, after computing the residues at $z = \pm \sqrt{\alpha}$ and $z =
0$, we get
\bea\label{5.32}
& A_N  =  \frac {(-2\pi^2)^N}{N} \frac
{d^N}{d\alpha^N} \mid_{\alpha = 4\pi^2} \sum^{\infty}_{s=0} \bigg [
\frac {5+2\eta_N}{(N+2s)!} -\frac{
  4\pi^2}{(N+2s+2)!} \bigg ] (-\alpha)^s \nonumber \\
& = \frac {(2\pi^2)^N}{N} \sum^{\infty}_{n=0} \frac {(N+n)!}{n!}
\bigg [ \frac {5+2\eta_N}{(3N+2n)!} -\frac {
  4\pi^2}{(3N+2n+2)!} \bigg ] (-4\pi^2)^n \, .
\eea
Again, the last series alternates in sign and decreases, which entails
\be\label{5.33}
A_N = (5+2\eta_{N})(2\pi^2)^N \frac {(N-1)!}{(3N)!} [1+O(\frac {1}{N} ] \, .
\ee
Finally, using  eqs. (\ref{5.13}), (\ref{5.30}), (\ref{5.33})
and (\ref{5.20}) we obtain
\be\label{5.34}
\frac{\rho^{(0)}_N}{N} = \frac{5+2\eta_N}{3N} \left[1+O\left(\frac{1}{N}\right)\right] =
\begin{array}{ll}
 \frac{2}{3} [1+O(\frac {1}{N})]  &\mbox {for bosons} \\
 \frac{1}{N} [1+O(\frac {1}{N})]  &\mbox {for fermions}
\end{array}
\ee
The same procedure applies for all integer values of $\beta$,
although the algebra
becomes quite involved. The general result for bosons (and for any
integer $\beta$) is:
\be\label{5.35}
\lim_{N \rightarrow \infty} \frac{\rho^{(0)}_N}{N}
= \frac{(\beta!)^4 [(3\beta+1)!]^2}{[(2\beta)!]^2[(2\beta+1)!]^3} \, .
\ee
Our method does not adapt straight
forwardly to the case of non-integer values of $\beta$, but there is clearly
no reason to expect a different outcome for such
intermediate values.
Therefore, the Onsager-Penrose criterion (\ref{5.12}) is {\bf not} met
for bosons,
and we reach the conclusion that Bose-Einstein condensation is
{\bf possible} in the bosonic version of the $N$-body model discussed in
Sect.3.

In the fermionic version, the result (\ref{5.34}) is not conclusive, as
explained after eq. (\ref{5.12}). It only points (not too surprisingly) to the
absence of quantum phase in the system.

\section{The  $B_N$ model in one dimension}

Subsequent to the seminal work
of Calogero and Sutherland for the $A_{N-1}$ system, the entire bound
state spectrum of the Calogero model was obtained
for $BC_N, D_N$ root systems \cite{op}.
It is then natural to enquire if
in our case, can one at least obtain the exact ground state and radial
 excitation spectrum in the $BC_N$ or
$D_N$ case? We now show that the answer to this question is in the affirmative.

Consider the $BC_N$ Hamiltonian,
\bea\label{4.1}
&& H  =  -{1\over 2}\sum^N_{i=1}{\partial^2\over\partial x^2_i} + V
\left(\sum^N_{i=1} x^2_i\right)+g \sum^{N-1}_{i=1} \bigg [{1\over (x_i-x_{i+1})^2} +
{1\over (x_i+x_{i+1})^2} \bigg ] \nonumber \\
&& - G \sum^{N-1}_{i=2} \bigg [\left({1\over x_{i-1} -x_i}
-{1\over x_{i-1} +x_i}\right)
\left({1\over x_i - x_{i+1}}+{1\over x_i +x_{i+1}}\right) \bigg ] \nonumber\\
&& + g_1 \sum_{i=1}^N {1\over x_i^2} \, ,
\eea
of which $B_N, C_N$ and $D_N$ are the special cases.
We again restrict our attention to the
sector of configuration space corresponding to a definite ordering of the
particles as given by eq. (\ref{5}).

We start with the ansatz
\be\label{4.2}
\psi = P_{2k} (x)\phi (r) \left(\prod^{N}_{i=1}
(x_i^2)^{\gamma /2} \right)
\prod^{N-1}_{i=1}(x^2_i-x^2_{i+1})^{\beta} \, ,
\ee
where $r^2 = \sum_{i=1}^{N} x_i^2$.
On substituting it in the Schr\"odinger equation for the
$B_N$-Hamiltonian (\ref{4.1}) we find that $\phi$ satisfies
\be\label{4.3a}
\Phi^{''} (r) + [N+4k -1+2N\gamma+4(N-1)\beta]{1\over r}\Phi' (r)
+2\left[ E-V(r)\right] \Phi (r) = 0 \, ,
\ee
provided $g$ and $G$ are again related to $\beta$ by eq. (\ref{2})
while $g_1$ is related to $\gamma$ by
\be
g_1 = \frac{\gamma}{2}(\gamma-1) \, .
\ee
Here, $P_{2k} (x)$ is
a homogeneous polynomial of degree
$2k$ ($k = 0,1,2,...$) in the particle-coordinates and satisfies
the generalized
Laplace equation
\be\label{4.4d}
 \bigg [ \sum^{N}_{i=1} {\partial^2 \over \partial x^2_i}
+2\gamma \sum_{i=1}^{N} {1\over x_i}{\partial \over \partial x_i}
+ 4\beta
 \sum^{N-1}_{i=1} {1\over (x_i^2 -x_{i+1}^2)}
\left(x_i {\partial \over\partial x_i}
- x_{i+1} {\partial
\over \partial x_{i+1}}\right) \bigg ] P_{2k}(x) = 0 \, .
\ee

Let us now specialize to the case of the oscillator potential, i.e., $V(r)
= {\omega^2\over 2} r^2$. In this case,  (\ref{4.3a}) is the well known
radial equation for the oscillator problem in
more than one dimension and its solution
is
\be\label{4.3b}
\Phi (r) = \exp ({-\omega r^2/2}) L^a_n (\omega r^2), \ n = 0,1,2,....
\ee
where $L^{a}_n(x)$ is the associated Laguerre polynomial while the energy
eigenvalues are given by
\be\label{4.3c}
E_{n} = \left[2n+2k+{N\over 2} +N\gamma + 2(N-1)\beta \right] \omega \, ,
\ee
with $a = {E\over \omega}-2n-1$.
The exact ground state is obtained from here when $n=k=0$.
The fact that $n=k=0$ gives the exact ground state energy
of the system can be easily shown {\it a la} $A_{N-1}$ case
by the method of supersymmetric quantum mechanics.
It may be noted that for large $N$, the
energy $E$ is proportional to $N$ so that like the
$A_{N-1}$ case, the $B_N$ model also
has a good thermodynamic limit. In contrast, notice that the
long-ranged $B_N$ Calogero model does not have a good thermodynamic limit.

Are there homogeneous polynomial solutions of eq. (\ref{4.4d}) of
degree $2k$ ($k \ge 1$)?
While we are unable to answer this question for any $k$, at least for
small values of $k$ ($k>0$)
there does not seem to be any solution to eq. (\ref{4.4d}). For
example, we have failed to find any polynomial solution of degree 2,4 and 6.
Thus it appears that unlike the $A_{N-1}$ case, in the $BC_N$ case one is only
able to obtain the ground state and radial excitations over it.

Proceeding in the same way,
the energy eigenvalues and eigenfunctions
in the case of the
Coulomb-like potential (\ref{4}) are
\be\label{4.9}
E = -{\alpha^2\over 2 \bigg [n+2k+{N-1\over 2}+N \gamma +2(N-1)\beta \bigg ]^2}
\ee
\be\label{4.10}
\Phi = e^{-\sqrt{2\mid E \mid r}} L^{b}_n(2\sqrt{2\mid E\mid} r)
\ee
where $b = N-2+4k+2N \gamma +4(N-1)\beta$. Again, so far we have been
able to obtain solutions only in case k=0.

As in Sec.II, in the $BC_N$ Hamiltonian (\ref{4.1}), all the particles
are not being treated on the same footing. Again, one possibility is to
add extra terms. Consider for example,
\be
H_1 = H + H' \, ,
\ee
where $H$ is as given by eq. (\ref{4.1}) while $H'$ has the form
\bea
H'&=& g\bigg [\frac{1}{(x_N -x_1)^2} +\frac{1}{(x_N+x_1)^2} \bigg ] \nonumber\\
&-& G\bigg [(\frac{1}{x_N -x_1} -\frac{1}{x_N+x_1})
(\frac{1}{x_1 -x_2}+\frac{1}{x_1 +x_2}) \nonumber \\
&+& \bigg (\frac{1}{x_{N-1} -x_N}-\frac{1}{x_{N-1}+x_N})
(\frac{1}{x_N -x_1} +\frac{1}{x_N +x_1}) \bigg ] \, .
\eea
One can now run through the arguments as given above and show that the
eigenstates for both the oscillator and Coulomb-like potentials
have the same form as given above except that in the term
multiplying $\beta$, $N-1$ gets replaced by $N$ at all places including in
the Laplace eq. (\ref{4.4d}). However, now we find that there are indeed
solutions to the Laplace eq. (\ref{4.4d}) (with $N-1$ replaced by $N$).
In particular, the solution for any $N (\ge 4$) and $k=4$ is given by
\be
P_{k=4} (x) = a\sum_{i=1}^{N} x_i^4 +b\sum_{i<j}^N x_i^2 x_j^2 \, ,
\ee
where
\be
\frac{b}{a} = -2\bigg [\frac{3+8\beta+2\gamma}{N-1+2(N-1)\gamma
+4(N-2)\beta} \bigg ] \, .
\ee
As in the $A_{N-1}$ case, we again find that even though the Laplace
eq. (\ref{4.4d}) is only invariant under cyclic permutations, the solution
is in fact invariant under the permutation of any two coordinates. It will
be interesting to try to find solutions for higher values of $k$ and study
the full degeneracy of the spectrum.

Besides these two,
one can obtain a part of the spectra including
the ground state for several other potentials
but we shall not discuss them here.

\section{$BC_N$ model in one dimension with periodic boundary condition}

Following the work of Sutherland \cite{sut} on the $A_{N-1}$ root system,
the exact ground state
as well as the excitation spectrum was also obtained in the case of the
$BC_N,D_N$ root systems \cite{op}.
It is then worth enquiring if, in
our case, one can also obtain the ground state and the excitation spectrum.
As a first step in that direction, we shall obtain the exact ground state of
the $BC_N$ model with periodic boundary condition.

The Hamiltonian for the $BC_N$ case is given by
\bea\label{9.1}
H & = & - {1\over 2} \sum^N_{i=1}{\partial^2\over\partial x^2_i}
+ g{\pi^2\over L^2} \sum^{N}_{i=1}
\bigg [{1\over \sin^2 {\pi\over L}(x_i-x_{i+1})}+
{1\over \sin^2 {\pi\over L}(x_i+x_{i+1})} \bigg ] \nonumber\\
 & + & g_1{\pi^2 \over L^2} \sum_i {1\over \sin^2 {\pi \over L}x_i}
 +  g_2{\pi^2 \over L^2} \sum_i {1\over \sin^2 {2\pi \over L}x_i}
  - G {\pi^2\over L^2} \sum^{N}_{i=1} \bigg [\cot {\pi \over L} (x_{i-1}
-x_i) \nonumber\\
 & - & \cot {\pi\over L}(x_{i-1}+x_i) \bigg ] \bigg [\cot {\pi\over L}
(x_i-x_{i+1})
+ \cot {\pi\over L} (x_i +x_{i+1}) \bigg ] \, .
\eea
We again restrict our attention to the
sector of the configuration space corresponding to a definite ordering of the
particles as given by eq. (\ref{2}).
For this case, we start with a trial wave function of the form
\be\label{9.3}
\Psi_0 = \prod^{N}_{i=1} \sin^{\gamma} \theta_i
\prod^N_{i=1} (\sin^2 2\theta_i)^{\gamma_1 /2}
\prod^N_{i=1} [\sin^2 (\theta_i-\theta_{i+1})]^{\beta /2}
\prod^N_{i=1} [\sin^2 (\theta_i+\theta_{i+1})]^{\beta 2} \, ,
\ee
($\theta_i = \pi x_i/L$) and substitute it in
the Schr\"odinger equation for the Hamiltonian (\ref{9.1}). We find that
it is indeed a solution provided $g$ and $G$ are again related to $\beta$
by eq. (\ref{2}) while $g_1,g_2$ are related to $\gamma,\gamma_1$ by
\be
g_1 = {\gamma \over 2}[\gamma+2\gamma_1 -1] \, , \ \
g_2 = 2\gamma_1 (\gamma_1 -1) \, .
\ee
The corresponding ground state energy turns out to be
\be\label{9.5}
E_0 = {N \pi^2\over 2L^2} (\gamma+\gamma_1+2\beta)^2 \, .
\ee
The fact that this is indeed the ground state energy can be easily proved
as in Secs. II and III.

\section{$N$-body problem in $D$-dimensions}

Having obtained some results for
the $N$-body problem (\ref{1}) in one dimension, we study
generalization to higher dimensions.
Let us consider the following model in $D$-dimensions :
\bea\label{3.1}
H & = & -{1\over 2} \sum^N_{i=1}\vec\nabla^2_i +g \sum^{N-1}_{i=1}{1\over
(\vec r_i-\vec r_{i+1})^2} \nonumber \\
  & - & G \sum^{N-1}_{i=2} {(\vec r_{i-1}-\vec r_i).
(\vec r_i-\vec r_{i+1})\over (\vec r_{i-1}-\vec r_i)^2
(\vec r_i-\vec r_{i+1})^2} + V\left(\sum^N_{i=1}\vec r^2_i\right) \, .
\eea
On using the ansatz,
\be\label{3.2}
\psi = \bigg (\prod^{N-1}_{i=1} \mid \vec r_i-\vec r_{1+1}\mid^{\beta} \bigg
)\phi (r) \, ,
\ \ r^2 = \sum^N_{i=1}\vec r^2_i \, ,
\ee
in the Schr\"odinger equation for the Hamiltonian (\ref{3.1}), it can be
shown that $\phi (r)$ satisfies
\be\label{3.3}
\phi^{''}(r) +[DN-1+2(N-1)\beta ]{1\over r}\phi'(r) +2(E-V(r))\phi (r)=0 \, ,
\ee
provided $g$ and $G$ are related to $\beta$ by
\be
g = \beta^2 +(D-2)\beta   \, , \ \ G = \beta^2 \, .
\ee
Equation (\ref{3.3})
is easily solved in the case of the oscillator potential (i.e., $V(r)
= {\omega^2\over 2} r^2)$ yielding the energy eigenstates as
\be\label{3.4}
\phi (r) = \exp \left(-{\omega \over 2} r^2\right) L^{b}_n (\omega r^2) \, ,
\ee
\be\label{3.5}
E_n = \left[2n+(N-1)\beta+{D N\over 2} \right] \omega \, .
\ee
Here $b = {E\over \omega}-2n-1$. It may be noted that as in all other
higher dimensional many-body problems, one has only obtained a part
of the energy eigenvalue spectrum which however includes the ground state.
In particular, the ground state energy eigenvalue and eigenfunction
is given by
\be\label{3.6}
E_0 = \left[(N-1)\beta +{DN\over 2} \right] \omega \, ,
\ee
\be\label{3.7}
\psi_0 = \exp \bigg (-{\omega \over 2}\sum^N_{i=1} r^2_i \bigg )
\prod^{N-1}_{i=1} \
\mid \vec r_i-\vec r_{i+1}\mid^{\beta} \, .
\ee
As expected, for $D$ = 1 these results go over to those obtained
in Sec. II. The fact that this is indeed the ground state energy can be
easily proved by using again a supersymmetric formulation \cite{cks}.

At this point it is worth asking if the probability distribution for $N$
particles (at least for some $D(> 1))$ can be mapped to some known random
matrix
ensemble ? In this context we recall that in the case of the
Calogero-type model, it has been shown that in two space dimensions
$\mid\psi_0\mid^2$ can be mapped to complex random matrix \cite{kr}.
Using this identification one was able to calculate all the correlation
functions of the corresponding many-body theory and show the
absence of long-range order but the presence of an off-diagonal
long-range order in that theory. Unfortunately, so far as
we are aware of, answer to this question is unknown in this particular
case. We hope that at least in the case of two space dimensions,
where $\mid\psi_0\mid^2$
for  our model is given by
\be\label{3.8}
|\psi_0 (z_i)|^2 = C \exp \left(-\omega \sum^{N}_{i=1} |z_i|^2\right)
\prod^{N-1}_{i=1}
|z_i - z_{i+1}|^{2\beta} \, ,
\ee
$\mid\psi_0\mid^2$ can
be mapped to some variant of the short-range Dyson model.

Finally, we observe that the ground state and a class of excited states can
also be obtained in $D$-dimensions in case the oscillator potential is
replaced by the $N$-body Coulomb-like potential $V(r) = -{\alpha/
{\sqrt{\sum {\bf r}^2_{i}}}}$, because the resulting
equation (\ref{3.3})
is essentially the radial equation for the Coulomb potential. In
particular, the energy eigenvalues and eigenfunctions are given by
\be\label{3.9}
E_n = - {\alpha^2 \over 2 \bigg [n + {DN-1\over 2} +(N-1)\beta \bigg ]^2} \, ,
\ee
\be\label{3.10}
\psi_n = \exp (-\sqrt{2|E|}r) L_n^{b'} (2\sqrt{2|E|r})
\bigg (\prod_{i=1}^{N-1} |{\bf r}_i - {\bf r}_{i+1}|^{\beta} \bigg ) \, ,
\ee
where $b' = DN-2+2(N-1)\beta$. It may again be noted that whereas the ground state energy is linear in $\beta$ in the oscillator case, it is not so in the case
of the Coulomb-like $N$-body potential.

\section{Short-range model in two dimensions with novel correlations}

Few  years back, Murthy et al. \cite{mu} considered a model in two
dimensions with two-body and three-body long-ranged interactions and
obtained the exact ground state and a class of excited states. The
interesting feature of this model was that all these states had a built-in
novel correlation of the form $\mid X_{ij}\mid^g$ where
\be\label{7.1}
X_{ij} = x_iy_j-x_jy_i \, .
\ee
It is then natural to enquire if one can construct a model in two dimensions
and obtain ground and few excited states of the system all of which would
have a built-in short-range correlation of the form
\be\label{7.2}
X_{j,j+1} = x_j y_{j+1}-y_j x_{j+1} \, .
\ee
We now show that this is indeed possible. Let us consider the following
Hamiltonian
\be\label{7.3}
H = -{1\over 2}\sum^N_{i=1} {\vec \nabla}^2_i
+{\omega^2 \over 2}\sum^N_{i=1} {\vec r}^2_i+
g \sum^{N-1}_{i=1} {{\vec r}^2_i+{\vec r}^2_{i+1} \over X^2_{i,i+1}} - G
 \sum^{N-1}_{i=2} {{\vec r}_{i-1} \cdot {\vec r}_{i+1}
\over X_{i-1,i} X_{i,i+1}}
\ee
where $X_{i,i+1}$ is as given by eq. (\ref{7.2}). We start with the
ansatz
\be\label{7.4}
\psi(x_i,y_i) = \left[ \prod^{N-1}_{i=1} X^{\beta}_{i,i+1} \right]
\exp \left({-{\omega \over 2} \sum_{i} {\vec r}^2_i}\right)
\phi (x_i,y_i) \, .
\ee
On substituting the ansatz in the Schr\"odinger equation
$H\psi = E\psi$, one finds that $\phi$
satisfies the equation
\bea\label{7.5}
\bigg [&-&{1\over 2} \sum^N_{i=1} {\vec \nabla}^2_i
 + \omega \sum^N_{i=1} {\vec r}_i \dot {\vec \nabla}_i
+\beta \sum^{N-1}_{i=1} {1\over X_{i,i+1}}
\bigg (x_{i+1}{\partial\over\partial y_i}
-y_{i+1} {\partial\over\partial x_i} \nonumber\\
&+& y_i {\partial\over\partial x_{i+1}}
-x_i {\partial\over\partial y_{i+1}} \bigg ) \bigg ] \phi
= \bigg (E-[N+2(N-1)\beta ]\omega \bigg )\phi \, ,
\eea
provided $g$ and $G$ are related by (2). It is interesting to note that
even though we are considering the novel correlation model
in two-dimensions, the relationship between $g$ and $G$ is as in the
case of our one-dimensional model. We do not know if this has
any deep significance.

We conclude from here
that $\psi$, as given by eq. (\ref{7.4}), with
$\phi$ being a constant is the ground state of the system with the
corresponding ground state energy being
\be\label{7.7}
E_0 = [N+2(N-1)\beta] \omega \, .
\ee
Let us remark that, like the relationship between
coupling constants,
the ground state energy too has essentially
the same form as that of the one-dimensional short-range
$A_{N-1}$ model as given by eq. (\ref{2.8}).
That one has indeed obtained the ground state can be proved as before.

As in other many-body problems in two and higher dimensions, we are unable
to find the complete excited-state spectrum. However, a class of excited
states can be obtained from (\ref{7.5}). To that end we introduce the
complex coordinates
\be\label{7.8}
z = x+iy, z^{*} = x-iy, \partial\equiv{\partial\over\partial z} = {1\over 2}
\left({\partial\over \partial x}
-i{\partial\over\partial y}\right), \partial^{*}\equiv
{\partial\over\partial z^{*}} = {1\over 2}
\left({\partial\over \partial x} +i{\partial\over\partial y}\right) \, .
\ee
In terms of these coordinates, the differential eq. (\ref{7.5}) takes the form
\bea\label{7.9}
\bigg [-2\sum^N_{i=1}\partial_i\partial_i^{*}
& + & 2\beta\sum_{i=1}^{N-1} {\bigg (z_{i+1}\partial_i-z_i\partial_{i+1}+
z^{*}_i\partial^*_{i+1}-z^{*}_{i+1}\partial^*_i \bigg )
\over \bigg (z_i z^{*}_{i+1}-z^{*}_i z_{i+1}\bigg )} \nonumber\\
& + & \omega \sum^N_{i=1}
\bigg (z_i\partial_i+z_i^{*}\partial_i^{*} \bigg ) -(E-E_0) \bigg ] \phi = 0.
\eea
Now it is readily proved shown that the Hamiltonian H commutes with the total
angular momentum
operator $L = \sum_{i=1}^{N} (z_i \partial_i - z_i^{*} \partial_{i}^{*}),$
so that one can classify solutions according to their angular momentum:
$L\phi = l\phi$.

On defining $t =\omega \sum_i z_i z^*_i$ and
let $\phi\equiv\phi(t)$ it is easily
shown that $\phi(t)$ satisfies
\be\label{7.10}
t\phi^{''}(t)+\left[{E_0\over \omega}-t\right]\phi' (t)
+\left({E-E_0 \over 2\omega}\right)\phi (t) = 0 \, ,
\ee
where $E_0$ is as given by eq. (\ref{7.7}).
Hence the allowed solutions with $l$ = 0 are
\be\label{7.11}
E = E_0 + 2n \omega, \ \phi(t) = L_{n}^{{E_0\over \omega}-1} (t) \, .
\ee
Solutions with angular momentum $l > 0$ or $ l < 0$ can similarly be obtained by
introducing $t_z = \omega \sum_i z_i^2$ or
$t_{z^{*}} = \omega \sum_i (z^{*}_i)^2$.
For example, let $\phi
= \phi(t_z)$. Then eq. (\ref{7.9}) reduces to
\be\label{7.12}
2\omega t_z \frac{d\phi}{dt_z} = (E-E_0)\phi \, .
\ee
This is the well known Euler equation whose solutions are just monomials in
$t_z$. The solution is given by $\phi (t_z) = t_z^m (m > 0)$,
and hence the angular
momentum $l =2m$ while the energy eigenvalues are $E = E_0 +2m\omega
= E_0 +l\omega$.
Further, we can combine these
solutions with the $l = 0$ solutions obtained above and obtain a tower of
excited states. For example, let us define
$\phi(z_i,z_i^{*}) = \phi_1 (t) \phi_2 (t_z)$, where $\phi_1$ is a solution
with $l=0$, while $\phi_2$ is the solution with $l > 0$. On using
$\phi_2 (t_z) = t_z^m$ it is easily shown that $\phi_1$ again satisfies a
confluent hypergeometric equation,
\be\label{7.13}
t\phi_1^{''}(t)+\left[{E_0\over \omega}+2m-t\right]\phi_1' (t)
+\left({E-E_0 \over 2\omega}+m \right)\phi_1 (t) = 0 \, .
\ee
Hence the energy eigenvalues are given by $E-E_0 = (2n_r +2m)\omega$. One may
repeat the procedure to obtain exact solutions for a tower of states with
$l < 0$.
\section{Summary}

In this paper we have discussed an $N$-body problem in one dimension and
presented its exact ground state on a circle and most likely the entire
spectrum on a real line.
There are several similarities as well as differences between the model
discussed here and Calogero-Sutherland (CSM) type of models and it might
be worthwhile to compare the salient features of the two.

\begin{enumerate}
\item Whereas in CSM the interaction is between all neighbours, in our case
the interaction is only between nearest and next-to-nearest neighbours.
Note however that in both the cases it is an inverse square interaction.
\item Whereas in CSM (in one dimension) there is only two-body interaction,
both two- and three-body interactions are required in our model
for partial (or possibly exact) solvability on a real line.
\item Whereas the complete bound state spectrum is obtained in the
Sutherland model (periodic potential) or if
there is external harmonic or Coulomb-like $N$-body potential as given by
eq. (\ref{4}) and in the case of both $A_{N-1}$ and $BC_N$ root systems,
it is not clear if this is so in our case even though it is likely
that this may be so in the $A_{N-1}$ case.
\item Whereas our system, both on a line and on a circle,
has  good thermodynamic limit (i.e. $E/N$ is finite
for large $N$), CSM does not have good thermodynamic limit in either case
and $E/N$ diverges
like $N$ for large $N$.
\item In both the cases,
the norm of the ground state wavefunction can be mapped
to the joint probability density function of the eigenvalues of some random
matrix. Using this correspondence, in both the cases, one is able to calculate
one- and two-point functions. However, whereas in the CSM this is possible
only at three values of the coupling (corresponding to orthogonal, unitary or
simplictic random matrices), in our case the correlation functions can be
computed analytically for any integral or half-integral values of the coupling
while numerically it can be done for any positive $\beta$.
\item In the CSM case with an external potential of the form
\be
V\left(\sum_i x_i^2\right) = A\sum_{i=1}^N x_i^2 +B\left(\sum_i x_i^2\right)^2
+ C\left(\sum_i x_i^2\right)^3
\ee
it has been shown \cite{jk1} that the
norm of the ground state wave function can
be mapped to a random matrix corresponding to branched polymers. It is not
known if a similar mapping is possible in our case.
\item A multi-species generalization of CSM has been done \cite{rs},
it is not clear if a similar generalization is possible in our case or not.
\item Generalization to $D$-dimensions ($D>1$)
is possible in CSM
as well as in our model and in both the
cases one is able to obtain only a partial spectrum
including the ground state. In both the cases, both two-
and three-body interactions are required.
Whereas our system has a good thermodynamic limit in any dimension $D$,
the CSM does not have a
good thermodynamic limit in any dimension.
However, whereas the norm of the ground state
wave function can be mapped to complex random matrices in the CSM case in
two dimensions \cite{kr},
no such mapping has so far been possible in our case for $D > 1$.
\item Model with novel correlations is possible in two dimensions in both
the cases \cite{mu} but unlike CSM, our system has a good thermodynamic
limit.
\item In the CSM, it has been possible to obtain the entire spectrum
algebraically by using supersymmetry and shape invariance \cite{gks}.
It would be nice if similar thing can also be done in our model. Further,
in the CSM, one has also written down the supersymmetric version of
the model \cite{fm}. It would be worth
enquiring if a similar thing can also be done
in our model.
\item In the CSM type models, one knows the various exactly solvable problems
in which the $N$-particles
interact pairwise by two body interaction \cite{cal}.
The question one would like to ask in our context is: what are the various
exactly solvable problems in one dimension in which the $N$ particles
have only nearest- and next-to-nearest neighbour interactions?
\item In the CSM, not only one- and two-point but even n-point correlation
functions are known. It would be nice if the same is also
possible in the present context.
\item {\it A la}  Haldane-Shastry spin models \cite{hs}, can we also construct
spin models in the context of our model?
\item Unlike CSM, in our case the off-diagonal long-range order is
nonzero in the bosonic version of the many-body theory in one dimension.
Note however that the off-diagonal long-range order is nonzero in the CSM in
two dimensions.
\end{enumerate}

\centerline {\bf Appendix }
\vskip .5 true cm
\begin{enumerate}
\item {\bf  Proof of the representation (\ref{5.17}) of $C_N$}

By construction, the square of the wave function (\ref{5.2}) is a symmetrical
function of all its arguments, so that we can write eq. (\ref{5.7}) as well:
\be\label{A1}
C_N = N! \int^1_0 dx_1 \int^{x_1}_0 dx_2 ... \int^{x_{N-1}}_0 dx_N
  \mid\psi_N (x_1,...,x_N)\mid^2 \, ,
\ee
where the particle coordinates are now properly ordered. We are thus
allowed to substitute $\phi_N$ for $\psi_N$ in (\ref{A1}) and obtain from
eqs. (\ref{5.3}) and (\ref{5.15})
\be\label{A2}
 C_N = N! \int^1_0 dx_1\int^{x_1}_0 dx_2 ... \int^{x_{N-1}}_0 dx_N
 \prod^N_{n=1} S(x_n - x_{n+1})^2 \, .
\ee
Changing the integration variables $(x_1,x_2,...,x_N)$ to
$(\triangle_1,\triangle_2,...,
\triangle_{N-1},x_N)$, where
\be\label{A3}
 \triangle_n = x_n - x_{n+1} \, ; \ \ (n=1,..., N-1) \, ,
\ee
one easily gets
\bea\label{A4}
 C_N & = & N! \int^1_0
d\triangle_1 \int^{1-\triangle_1}_0 d\triangle_2 ...
\int^{1-\triangle_1-...-\triangle_{N-2}}_0
d\triangle_{N-1} \ X \nonumber \\
     & X & \int^{1-\sum^{N-1}_{p=1} \triangle_p}_0 dx_N
\prod^{N-1}_{n=1} S(\triangle_n)^2 S(x_N-x_1)^2 \, .
\eea
Since $x_N-x_1 = - \sum^{N-1}_{p=1} \triangle_p$ is in fact independent
of $x_N$ in the new set of variables, eq. (\ref{A4}) becomes, using also
$S(-x) = S(1-x)$:
\bea\label{A5}
C_N & = & N! \int^1_0
d\triangle_1 \int^{1-\triangle_1}_0
d\triangle_2 ... \int^{1-\triangle_1-...-\triangle_{N-2}}_0
d\triangle_{N-1} \ X \nonumber \\
    & X & (1-\sum^{N-1}_{p=1}
\triangle_p)
\prod^{N-1}_{n=1} S(\triangle_n)^2
S(1-\sum^{N-1}_{p=1}\triangle_p)^2  \, .
\eea
It is now convenient to introduce the extra variable
\be\label{A6}
 \triangle_N = 1 - \sum^{N-1}_{p=1} \triangle_p \, ,
\ee
and to recast eq. (\ref{A5}) in the form
\bea\label{A7}
 C_N & = & N! \int^1_0
d\triangle_1 \int^1_0 d\triangle_2 ... \int^1_0
d\triangle_{N-1} \int^1_0 d\triangle_N
\delta (1-\sum^{N}_{p=1} \triangle_p) \ X \nonumber \\
     & X & \triangle_N
\prod^{N}_{n=1} S(\triangle_N)^2 \nonumber \\
     & = & N! \int^1_0
d\triangle_1 ... \int^1_0 d\triangle_N
\delta  (1-\sum^N_{p=1} \triangle_p) \frac {1}{N}
\sum^N_{m=1}\triangle_m \prod^N_{n=1} S(\triangle_n)^2 \nonumber \\
     & = & (N-1)! \int^1_0
d\triangle_1 ... \int^1_0 d\triangle_N
\delta  (1-\sum^N_{p=1} \triangle_p)
\prod^N_{n=1} S(\triangle_n)^2  \, .
\eea
In the second equality, we have used the fact that, apart from the
factor $\triangle_N$, the integrand and the integration range are
completely symmetrical in the variables
$(\triangle_1,...,\triangle_N)$. Finally, the integration over these
variables factorizes after introducing the representation
\be\label{A8}
\delta (1-\sum^N_{p=1}\triangle_p) = \frac{1}{2\pi}
\int^{\infty}_{-\infty} dx e^{-ix(1-\sum^N_{p=1}\triangle_p)} \, ,
\ee
and interchanging the x- and $\triangle -$integrations. This produces
eq. (\ref{5.17}).

\item {\bf Proof of the representation (\ref{5.18}) of $A_N$}

Proceeding along the same lines, we first put the expression (\ref{5.14}) of
$A_N$ in the form
\bea\label{A9}
 A_N & = & (N-1)! \int^1_0 dx_1 \int^{x_1}_0 dx_2 ... \int^{x_{N-2}}_0
dx_{N-1} \phi_N (x_1,...,x_{N-1},0) \ X \nonumber \\
     & X & R_N(x_1,...,x_{N-1}) \, ,
\eea
where
\bea\label{A10}
&& R_N(x_1,...,x_{N-1}) = \int^{x_{N-1}}_0
dx \phi_N (x_1,...,x_{N-1},x) \nonumber \\
&& \pm \int^{x_{N-2}}_{x_{N-1}}
dx \phi_N (x_1,...,x,x_{N-1})
+ ... + \int^1_{x_1} dx \phi_N
(x,x_1,...,x_{N-1}) \nonumber \\
&& = \int^{x_{N-1}}_0
dx \phi_N (x_1,...,x_{N-1},x)
+\int^1_{x_1}
dx \phi_N (x,x_1,...,x_{N-1}) \nonumber \\
&& +\sum^{N-2}_{p=1} \nu_p \int^{x_p}_{x_{p+1}}
dx \phi_N (x_1,...,x_p,x,x_{p+1},...,x_N) \, .
\eea
Here, $\nu_p = 1 (\nu_p=(-1)^{p})$ for bosons (fermions) and we have
used the restriction to odd $N$ in the second case. Thanks to the
periodicity and the cyclic symmetry of $\phi_N$ , the first two terms in
the last expression above can be collected to give

$$ \int^{x_{N-1}}_{x_{1}-1} dx \phi_N (x,x_1,...,x_{N-1}) \, . $$

Hence $R_N$ becomes (with \ $x_N = x_1 -1$)
\bea\label{A11}
&& R_N (x_1,...,x_{N-1}) = \sum^{N-1}_{p=1}\nu_p \int^{x_p}_{x_{p+1}} dx
\phi_N (x_1,...,x_p,x,x_{p+1},...,x_{N-1}) \, \nonumber \\
&& =  \sum^{N-1}_{p=1} \nu_p \prod^{N-1}_{n=1} S(x_{n} - x_{n+1})
\int^{x_p}_{x_{p+1}} dx S(x-x_{p+1}) S(x_p-x) \, , \; \;  (n \ne p) \nonumber\\
&& = \sum^{N-1}_{p=1} \nu_p \sum^{N-1}_{n=1} S(x_n-x_{n+1})
 S_2 (x_p - x_{p+1}) \, , \; \; \; (n \ne p)
\eea
according to the definition (\ref{5.16}). We also have:
\be\label{A12}
\phi_N (x_1,...,x_{N-1},0) = \prod^{N-2}_{m=1} S(x_m-x_{m+1})
S(x_{N-1}) S(x_1) \, .
\ee
Inserting eqs. (\ref{A11}) and (\ref{A12}) in  eq. (\ref{A9}) and
introducing as before
the new integration variables $\triangle_n \equiv x_{n} - x_{n+1}$ \ ($n=1,...,
N-2)$ and $x_{N-1}$, we obtain
\bea\label{A13}
A_N & = & (N-1)! \int^1_0 d\triangle_1 \int^{1-\triangle_1}_0 d\triangle_2
... \int^{1-\triangle_1-...-\triangle_{N-3}}_0 d\triangle_{N-2} \ X \nonumber \\
    & X & \int^{\triangle_{N-1}}_0 dx_{N-1}
\prod^{N-2}_{m=1}
S(\triangle_m) S(x_{N-1}) S(x_{N-1}-\triangle_{N-1}) \ X \nonumber \\
    & X & \sum^{N-1}_{p=1} \nu_p \prod^{N-1}_{n=1} S(\triangle_n)
S_2(\triangle_p) \, ; \; \; \; (n \ne p)
\eea
where $\triangle_{N-1}=1-\sum^{N-2}_{p=1}\triangle_p$. The integration
over $x_{N-1}$ gives the factor $S_2(\triangle_{N-1})$ in place of
$S(x_{N-1}) S(x_{N-1}-\triangle_{N-1})$, so that
\bea\label{A14}
&& A_N = (N-1)! \int^1_0 d\triangle_1 ... \int^1_0
d\triangle_{N-1} \delta (1-\sum^{N-1}_{p=1}\triangle_p)
 \bigg [ \prod^{N-2}_{m=1} \ X \nonumber \\
&& X \; \;  S(\triangle_m)S_2(\triangle_{N-1})
\sum^{N-1}_{p=1}
\nu_{p} \prod^{N-1}_{n=1} S(\triangle_n)S_2(\triangle_p) \bigg ] \, ,
\; \; \;  (n \ne p) \, .
\eea
On taking into account the complete symmetry of the integration measure,
one finds that the square bracket in eq. (\ref{A14}) can be replaced by
\bea\label{A15}
 [...] & = & \prod^{N-2}_{m=1}
S(\triangle_m)^2 S_2(\triangle_{N-1})^2
+ \eta_N \prod^{N-3}_{m=1}
S(\triangle_m)^2 \ X \nonumber \\
       & X & [S(\triangle_{N-2})S_2(\triangle_{N-2})] [
S(\triangle_{N-1}) S_2(\triangle_{N-1})] \, ,
\eea
where $\eta_N$ is as defined in eq. (\ref{5.20}). Finally, one obtains
the factorization of the multiple integral in eq. (\ref{A14})
by using again the
representation (\ref{A8}) of the $\delta$ measure (with $(N-1)$ in place of
$N$). This entails eq. (\ref{5.18}).
\end{enumerate}

A last remark may be in order. Alternative, equivalent forms of the
representations (\ref{5.17}) and (\ref{5.18}) would be obtained
by relying on Fourier
expansions instead of Fourier integrals, that is by considering the
integrands in eqs. (\ref{A5}) and (\ref{A13}) not as functions with compact
supports $[0,1]^N \subset {\cal{R}}^N$, \ resp. \ $[0,1]^{N-1}
\subset {\cal{R}}^{N-1}$,
but as periodic functions (this would amount to modifying  eq. (\ref{A8})
accordingly). It turns out however that the resulting representations
of $C_N$ and $A_N$ (as Fourier series) are much less convenient for
the explicit or asymptotic evaluations of these quantities.

\noindent
{\bf Acknowledgements}

SRJ acknowledges the warm hospitality of the Institute of Physics,
Bhubaneswar where this work was initiated
while AK would like to thank the members of the Laboratoire de Physique
Math\'ematique of
Montpellier University for warm hospitality during his trip there as
a part of the Indo-French Collaboration Project 1501-1502.

\newpage

\noindent
{\large \bf Figure Legends}
\vskip 1.0 truecm

\noindent
{\bf Fig. 1} The two-point correlation function for four integer
values of $\beta $ (from left to rightmost are increasing values from 1
to 4) shows clearly an absence of long-range order.

\vskip 1.0 truecm

\noindent
{\bf Fig. 2} The two-point correlation function for some fractional
values of $\beta $ plotted alongwith $\beta $ equal to 1 and 2.
>From left to rightmost are increasing values from 1, 4/3, 3/2, 5/3, 2,
7/3, and 5/2. Thus, even for fractional values, there is no
long-range order.

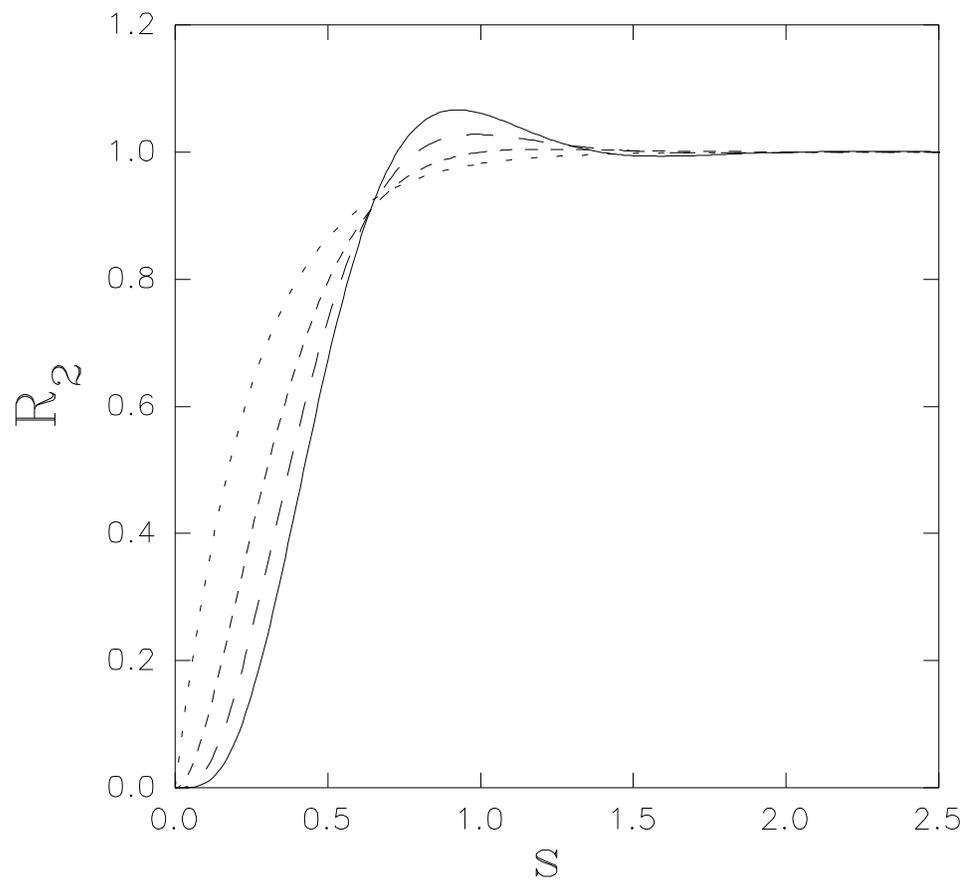
\end{document}